\documentstyle[12pt]{article}

\newcommand{\0}{{\bf 0}}
\newcommand{\1}{{\bf 1}}

\newcommand{\aaa}{{\cal A}}
\newcommand{\ba}{{\cal B}}
\newcommand{\be}[1]{\begin{equation}\label{#1}}
\newcommand{\bra}[1]{{\langle #1 |}}
\newcommand{\braket}[2]{{\langle #1 |\, #2\rangle}}
\newcommand{\bracket}[3]{{\langle #1 |#2|\, #3\rangle}}
\newcommand{\card}{{\rm card}}

\newcommand{\dd}{{d}}
\newcommand{\ddd}{{D}}
\newcommand{\degree}{{\rm deg}}

\newcommand{\ee}{\end{equation}}

\newcommand{\hhh}{{\cal H}}

\newcommand{\ket}[1]{{|\, #1\rangle}}
\newcommand{\ketbra}[2]{{|\, #1\rangle\langle #2|}}
\newcommand{\kkk}{{\cal K}}
\newcommand{\lll}{{\cal L}}

\newcommand{\mmm}{{\cal M}}
\newcommand{\om}{\omega}
\newcommand{\omg}{\Omega}
\newcommand{\ppp}{{P}} 
\newcommand{\rrr}{{\cal R}}

\newcommand{\spn}{{\rm span}}
\newcommand{\ttt}{{\cal T}}

\title{Differential structure of Greechie logics}

\author{Roman B. Breslav, Roman R.Zapatrin}
\date{}

\begin{document}

\maketitle
\begin{center}
   {\em A.A. Friedmann Laboratory for Theoretical Physics, \\
     SPb UEF, Griboyedova 30-32, 191023, St.Petersburg,
     Russia
    }

\end{center}

\begin{abstract}
A liaison between quantum logics and non-commutative
differential geometry is outlined: a class of quantum logics are
proved to possess the structure of discrete differential
manifolds. We show that the set of proper elements of an
arbitrary atomic Greechie logic is naturally endowed by Koszul's
differential calculus.
\end{abstract}

\section*{Introduction}

In this paper we explore a liaison between quantum logic and
non-commutative geometry. Namely, we show that there is a class
of quantum logics which carries a natural differential
structure.

It was established yet by Koszul (1960) that differential
calculus on smooth manifolds admits a purely algebraic
reformulation in terms of graded differential modules over
algebras of smooth functions. Several versions of
non-commutative geometry --- operator extensions of classical
(sometimes called commutative) geometry stem from Koszul's
formalism. Geometrical models based on finite-dimensional
algebras were an object of particular interest (see Baehr {\em
et al.}, 1995; Zapatrin, 1997) due to their possible relevance
to 'empirical quantum geometry'. This research resulted in the
formalism of {\em dicrete differential manifolds} - finite sets
whose algebra of functions is endowed with a 'differential
envelope'. It was proved that many classical geometrical
features (Dimakis and M\"uller-Hoissen 1998) survive in these
models.

Our paper is organized as follows. In section \ref{s1} we
associate (following Rota, 1968) with an arbitrary poset $\ppp$
a non-commutative algebra $\omg$ called incidence algebra of
$\ppp$. Then we show that if $\ppp$ is 'good enough' tha algebra
$\omg$ acquires some useful properties, for instance, becomes
graded. In section \ref{s2} we give a brief outline of Koszul's
(1960) formalism of the calculus of differentials and introduce
the general notion of discrete differential manifold. In section
\ref{s3} we introduce a class of posets, called differentiable,
which can be treated as discrete differential manifolds and,
finally, show that atomic Greechie logics are always
differentiable.

\section{Incidence algebras}\label{s1}

\subsection{Algebras of scalars in Dirac's notation}\label{ss11}

Let $P$ be a set. Denote by $\hhh$ the space of all finite
formal linear combinations fo elements of $P$ written as Dirac's
ket-vectors:

\be{e37c}
\hhh = \spn\{\ket{p}\}_{p\in P}
\ee

\noindent and by $\hhh^*$ the dual to $\hhh$ spanned on the
basis of bra-vectors:

\[
\hhh^* = \spn\{\bra{q}\}_{q\in P}
\]

\noindent such that

\be{e37}
\braket{p}{q} = \delta_{pq} =
\left\lbrace \begin{array}{rcl}
1 &,& \mbox{if } p=q \cr
0 && \mbox{otherwise}
\end{array} \right.
\ee

\medskip

Now consider the set of symbols $\ketbra{p}{p}$ for all $p\in P$
and its linear span

\be{e37a}
\aaa = \spn\{\ketbra{p}{p}\}_{p\in P}
\ee

\noindent and endow it with the operation of multiplication

\be{e37b}
\ketbra{p}{p}\cdot\ketbra{q}{q} =
\ket{p}\braket{p}{q}\bra{q} =
\left\lbrace \begin{array}{rcl}
\ketbra{p}{p} &,& \mbox{if } p=q \cr
0 && \mbox{otherwise}
\end{array} \right.
\ee

\noindent making $\aaa$ associative and commutative algebra.

We have defined $\aaa$ as algebra of formal symbols
(\ref{e37a}). However, the elements of $\aaa$ can be treated as
operators on both $\hhh$:

\[
(\ketbra{p}{p})\ket{q} := \ket{p}\braket{p}{q} =
\left\lbrace \begin{array}{rcl}
\ket{p} &,& \mbox{if } p=q \cr
0 && \mbox{otherwise}
\end{array} \right.
\]

\noindent and $\hhh^*$:

\[
\bra{q}(\ketbra{p}{p}) := \braket{q}{p}\bra{p} =
\left\lbrace \begin{array}{rcl}
\bra{p} &,& \mbox{if } p=q \cr
0 && \mbox{otherwise}
\end{array} \right.
\]

\subsection{Incidence algebras of posets and their moduli of
differentials}\label{ss12}

The notion of incidence algebra of a poset was in troduced by
Rota (1968) in a purely combinatorial context. Let $P$ be
a partially ordered set:  $P=(P,\le)$.  Take all ordered pairs
$p,q$ of elements of $P$, form the linear hull

\be{38a}
\omg = \spn\{\ketbra{p}{q}\}_{p\le q}
\ee

\noindent and extend the formula (\ref{e37b}) to define the
product in $\omg$:

\be{38}
\ketbra{p}{q} \cdot \ketbra{r}{s} =
\ket{p} \braket{q}{r} \bra{s} =
\braket{q}{r} \cdot \ketbra{p}{s} =
\left\lbrace \begin{array}{rcl}
\ketbra{p}{s} &,& \mbox{if } q=r \cr
0 && \mbox{otherwise}
\end{array} \right.
\ee

One may doubt in the correctness of this definition of the
product: who guarantees us that $\ketbra{p}{s}$ is still in
$\omg$ when $q=r$? But recall that $P$ is partially ordered:
$\ketbra{p}{q}\in \omg$, $\ketbra{q}{s}\in \omg$ means $p\le q$
and $q\le s$, therefore $p\le s$, that is why $\ketbra{p}{s} \in
\omg$. The obtained algebra $\omg$ with the product (\ref{38})
is called {\em incidence algebra} of the poset $P$.

The incidence algebra $\omg$ is obviously associative, but not
commutative in general. The algebra $\aaa$ of scalars is a
maximal commutative subalgebra of $\omg$.

Let us split $\omg$ considered {\em linear space} rather than
algebra into two subspaces

\[
\omg = \aaa \oplus \rrr
\]

\noindent and call

\[
\rrr = \spn \{ \ketbra{p}{q} \}_{p<q}
\]

\noindent {\em module of differentials} of the poset $P$. In
fact, it follows directly from (\ref{38}) that for any $a\in
\aaa$ and any $\om \in \rrr$ both $a\om$ and $\om a$ are in
$\rrr$. It also follows directly from (\ref{38}) that for any
$a,b\in \aaa$, $\om \in \rrr$

\[
(a\om)b = a(\om b)
\]

\noindent therefore the module of differentials $\rrr$ is always
$\aaa$-bimodule\footnote{Although $\aaa$ is commutative, $a\om
\neq \om a$ in general}.

\subsection{Incidence algebras of Jordan-H\"older
posets}\label{ss13}

Recall that a poset $P$ is said to satisfy the {\em
Jordan-H\"older condition} (Birkhoff, 1967) if for any ordered
pair $p,q\in P$, $p<q$ the lengths of all maximal chains
$p<r<\ldots <s<q$ are equal. In this case with every basic
element $\ketbra{p}{q}$ of $\omg$ the following nonnegative
integer can be associated:

\be{e40a}
\degree\ketbra{p}{q} = \mbox{'the length of a maximal chain between
$p$ and $q$'}
\ee

\noindent splitting $\omg$ into linear subspaces

\be{40}
\omg = \omg^0 \oplus \omg^1 \oplus \ldots
\ee

\noindent with

\[
\begin{array}{rcl}
\omg^0 &=& \spn\{\ketbra{p}{p}\} \: = \: \aaa \cr
\ldots & \ldots & \ldots \cr
\omg^n &=& \spn\{\ketbra{p}{q}\}_{\deg\ketbra{p}{q} = n} \cr
\ldots & \ldots & \ldots
\end{array}
\]

\noindent making $\omg$ graded algebra:

\[
\forall \om \in \omg^m, \om' \in \omg^n \qquad
\om\om' \in \omg^{m+n}
\]

\noindent and therefore making the module of differentials
$\rrr$ graded $\aaa$-bimodule:

\[
\rrr = \omg^1 \oplus \omg^2 \oplus \ldots
\]

\medskip

\section{Graded differential algebras} \label{s2}

An algebraic version of differential calculus on manifolds due
to Koszul (1960) is presented in this section. It admits
powerful generalizations which gave rise, in particular, to
non-commutative geometry (in Dubois-Violette (1988) version, see
Djemai (1995) for an outline).

Let $\mmm$ be a smooth manifold. Denote by $\aaa$ the algebra of
smooth functions on $\mmm$ and by $\ttt^*$ the cotangent bundle
over $\mmm$:

\[
\aaa = C^\infty (\mmm) \quad ; \quad \ttt = T^*(\mmm)
\]

The elements of the exterior product $\ttt^* \wedge \ldots
\wedge \ttt^*$ are called differential forms. Denote

\[
\begin{array}{rclcl}
\omg^0 &=& \aaa & \qquad & \mbox{the space of scalars} \cr
\omg^1 &=& \ttt^* & \qquad & \mbox{the space of 1-forms} \cr
\ldots & \ldots & \ldots & \ldots & \ldots \cr
\omg^n &=& \wedge_n \ttt^* & \qquad & \mbox{the space of
$n$-forms} \cr
\ldots & \ldots & \ldots & \ldots & \ldots
\end{array}
\]

\noindent and form the direct sum

\[
\omg = \omg^0 \oplus \omg^1 \oplus \ldots
\]

\noindent which is graded algebra with respect to the exterior
product $\wedge$ of differential forms:

\[
\forall \om \in \omg^m, \om' \in \omg^n \qquad
\om \wedge \om' \in \omg^{m+n}
\]

The module of differentials

\[
\rrr = \omg^1 \oplus \omg^2 \oplus \ldots
\]

\noindent is graded $\aaa$-bimodule. In classical differential
geometry $\forall a \in \aaa, \om \in \omg^n \quad a\om = \om
a$.

\medskip

An important operator defined on $\omg$ makes it {\em
differential calculus}. This is the Cartan differential $\ddd:
\omg \to \omg$ which has the following properties:

\be{e42}
\begin{array}{l}
\ddd(\omg^m) \subseteq \omg^{m+1} \cr
\ddd^2 = 0 \cr
\ddd \1 = 0 \quad \mbox{(where $\1$ is the unit element of the
algebra $\aaa$)}\cr
\forall\om \in \omg^m, \om' \in \omg^n \quad
\ddd(\om\wedge \om') = \ddd\om\wedge\om' + (-1)^m\om\wedge\om'
\end{array}
\ee

\medskip

In general, a graded algebra $\omg$ built from an algebra $\aaa$
endowed with and operator $\ddd$ satisfying (\ref{e42}) is
called {\em differential calculus} over the algebra $\aaa$.

\medskip

The entire differential structure of the manifold $\mmm$ is
captured in its differential algebra $\omg$. For instance, the
space of vector fields is the dual to $\omg^1$.

Differential calculi over finite-dimensional commutative
algebras were thoroughly studied in the last decade. As a result
of this research an analog of (pseudo-) Riemannian geometry on
finite and discrete sets was built (see Baehr {\em et al.,}
1995; Dimakis and M\"uller-Hoissen, 1998). The triples

\[
(P, \omg(P), \ddd)
\]

\noindent are referred to as {\em discrete differential spaces},
where $P$ is a set (at most countable), $(\omg(P), \ddd)$ is a
graded differential algebra over the algebra $\aaa$ of scalars
on $P$.

\medskip

In our paper we show that if we take a Greechie logic $\lll$
then the order structure on $\lll$ induces (in unambiguous way!)
a differential calculus making it discrete differential space.

\medskip

\section{Differential calculi on Greechie logics}\label{s3}

\subsection{Differentiable posets}\label{ss31}

A poset $\ppp$ is said to be {\em differentiable} if it

\begin{itemize}
\item possesses Jordan-H\"older property (see section
\ref{ss13})
\item admits an operator $\dd$ on the space $\hhh$ of scalars on
$\ppp$, $\dd:\hhh \to \hhh$ such that
\end{itemize}

\be{e49}
\begin{array}{l}
\dd^2 = 0 \cr
\deg\ketbra{dp}{q} = \deg\ketbra{p}{q} + 1
\end{array}
\ee

\noindent where $\deg$ is the degree of elements of the
incidence algebra $\omg(\ppp)$ defined like in
(\ref{e40a}): let $\dd \ket{p} = \sum\epsilon_s \ket{s}$, then
$\ketbra{\dd p}{q}$ denotes the sum of only such $\epsilon_s
\ket{s}$ for which $\ketbra{s}{q} \in \omg$.

When $\ppp$ is a differential poset we are always ina position
to define the operator $\ddd:\omg \to \omg$ as follows:

\be{49a}
\ddd(\ketbra{p}{q}) :=
\ketbra{\dd p}{q} - (-1)^{\deg\ketbra{p}{q}}\ketbra{p}{q\dd}
\ee

\noindent where $\bra{q\dd}$ the action of the adjoint to $\dd$
operator on bra-vectors: see (\ref{e16a}) below.

We claim that the operator $\ddd$ has the properties
of Cartan differential. Verify the conditions (\ref{e42}) for
$\ddd$. Let $\ketbra{p}{q} \in \omg^m$ then

\[
\deg\ketbra{dp}{q} = \deg\ketbra{p}{q} + 1
\]

\noindent according to (\ref{e49}). To verify the second
condition calculate the value of $\ddd \1$ on an arbitrary basic
vector $\ket{r}$ of $\hhh$. Since $\1 = \sum\ketbra{p}{p}$ we
have

\[
\ddd \1 \ket{r} =
\left( \sum_{p\in \ppp}\ddd \ketbra{p}{p}\right)\ket{r} =
\sum_{p\in \ppp}\ket{\dd p}\braket{p}{r} -
\sum_{p\in \ppp}\ket{p}\braket{p\dd}{r} =
\]
\[
\ket{\dd r} -
\left(\sum_{p\in \ppp}\ketbra{p}{p}\right)\ket{\dd r} =
\ket{\dd r} - \1 \ket{\dd r} = 0
\]

To verify the third condition (\ref{e42}) let $\om =
\ketbra{p}{q}$ with $\deg\ketbra{p}{q}=m$ and let $\om' =
\ketbra{r}{s}$ with $\deg\ketbra{r}{s}=n$. Then

\[
\ddd(\om\om') =
\braket{q}{r}\ddd\ketbra{p}{s} =
\braket{q}{r}(\ketbra{\dd p}{s} -
(-1)^{m+n}\ketbra{p}{s \dd})
\]

\noindent On the other hand we have

\[
\ddd\om\cdot\om' =
(\ketbra{\dd p}{q} - (-1)^m\ketbra{p}{q \dd})\ketbra{r}{s} =
\braket{q}{r}\ketbra{\dd p}{s} -
\bracket{q}{\dd}{r}(-1)^m\ketbra{p}{s}
\]

\[
\om\cdot\ddd\om' =
\ketbra{p}{q}(\ketbra{\dd r}{s} -
(-1)^n\braket{q}{r}\ketbra{p}{s\dd})
\]

\noindent Therefore

\[
\ddd\om\cdot\om' + (-1)^m \om\cdot\ddd\om' =
\braket{q}{r}(\ketbra{\dd p}{s} -
(-1)^{m+n}\ketbra{p}{s \dd}) =
\ddd(\om\om')
\]

So, we conclude that any differential poset becomes a discrete
differential manifold whenever a border operator $\dd$
(\ref{e49}) is specified.

\subsection{Differential structure on simplicial complexes}

A {\em simplicial complex} $\kkk = (\kkk, V)$ is a collection
$\kkk$ of of nonempty subsets (called {\em simplices}) of a set
$V$ (called the set of {\em vertices} of $\kkk$) such that

\[
\forall s,s' \subseteq V \qquad s\in \kkk \mbox{ and }
s'\subseteq s \mbox{   imply  } s'\in \kkk
\]

In particular, a simplex is a complex consisting of {\em all}
nonempty subsets of the set of its vertices.

Any simplicial complex $\kkk$ consists of simplices which
are, in turn, sets. That is why $\kkk$ is partially ordered by
set inclusion. With any simplex $p$ a positive integer $\# p$ is
associated from the cardinality of $p$ considered set:

\[
\# p = \card\{ v\in V :\: v \in p\} - 1
\]

\noindent Consider the incidence algebra $\omg = \omg(\kkk)$
(section \ref{ss12}) of the complex $\kkk$. With any
$\ketbra{s}{t} \in \omg$ we associate

\be{50}
\deg \, \ketbra{s}{t} := \#s - \#t
\ee

\noindent making the algebra $\omg$ graded:

\[
\omg = \omg^0 \oplus \omg^1 \oplus \ldots
\]

In any simplicial complex the border operator $\dd$ is defined

\be{46}
\dd \ket{p} = \sum \epsilon_s \ket{s}
\ee

\noindent with $\epsilon_s = \pm 1$ such that

\be{46a}
\begin{array}{rl}
\dd^2 = 0 & \cr
\forall v\in V & \dd \ket{v} = 0 \cr
\mbox{if } \# p = m & \mbox{ then for any } s \mbox{ from
(\ref{46})}\quad \# s = m-1
\end{array}
\ee

The adjoint to $\dd$, called {\em coborder operator} acts in
$\hhh^*$. Due to Dirac's notation we may use the same symbol
$\dd$ for both border and coborder operators with no confusion:

\[
\dd: \bra{p} \mapsto \bra{p\dd}
\]

\noindent so that

\be{e16a}
\braket{p}{\dd q} = \braket{p\dd}{q} = \bracket{p}{\dd}{q}
\ee

Let us verify the conditions (\ref{e49}) for an arbitrary
simplicial complex $\kkk$. The first condition (\ref{e49})
follows from the first condition (\ref{46a}) and the second
condition (\ref{e49}) follows from (\ref{50}) and
the third condition of (\ref{46a}), therefore

\begin{itemize}
\item any simplicial complex $\kkk$ is differentiable poset
\item the border operator on $\kkk$ makes the set of its
simplices discrete differential manifold
\end{itemize}

\subsection{The differential structure on atomic Greechie
logics}\label{ss33}

An atomic $\sigma$-orthomodular poset $\lll$ is called {\em
atomic Greechie logic} if it can be represented as a union of
almost disjoint Boolean algebras:

\be{51}
\begin{array}{rcl}
\lll &=& \cup_i \ba_i \cr
&& \cr
\ba_i \cap \ba_j &=&
    \left[ \begin{array}{l}
    \{\0,\1\} \cr
    \{\0,\1,\, v, v'\}
    \end{array}\right.
\end{array}
\ee

\noindent where $v$ is an atom of $\lll$.

\medskip

Let $\lll$ be a Greechie logic with the set of atoms $V$. In
this section we show that the poset $\ppp$ of {\em proper}
elements of $\lll$:

\be{52}
\ppp := \lll \setminus \{\0,\1\}
\ee

\noindent is differentiable and build the border operator on
$\ppp$ making it discrete differential manifold.

Let us build the simplicial complex $\kkk = (\kkk,V)$ starting
from the decomposition (\ref{51}) of $\lll$. The set $V$ of
atoms of $\lll$ will be the set of vertices of $\kkk$. A
non-empty subset $s\subseteq V$ will be a simplex of $\kkk$
whenever $s$ is a {\em proper} (sic!) subset of atoms of a block
$\ba_i$ of $\lll$:

\[
\kkk :=
\{ s\subseteq V :\:
\exists \ba_i \: s\subset V(\ba_i)\,,\: s\neq \0, V(\ba_i)\} 
\]

\noindent where $V(\ba_i)$ is the set of atoms of the block
$\ba_i$.
The poset $\ppp$ is Jordan-H\"older. To prove it, let
$p,q\in \ppp, p < q$, then (since they are proper elements of
$\lll$) there is a unique block $\ba_i$ from (\ref{51}) which
contains them. We put

\be{54}
\deg \ketbra{p}{q} := \#_i q - \#_i p
\ee

\noindent like in (\ref{50}) with

\[
\#_i p = \card \{v \in \ba_i :\: v\le p\}
\]

\medskip

With every simplex $s\in \kkk$ an element of the poset $\ppp$
(\ref{52}) can be associated. Take the mapping $f$ from $\kkk$
to $\lll$

\be{52a}
f(s) := \vee_\lll \{v\in V :\: v\in s\}
\ee

\noindent which is surjective (since any element of $\ppp$ is
contained in a block and thus can be expressed as a join of
atoms) but not injective (since an element of $\ppp$ can belong
to more than one block).

Extend the mapping (\ref{52a}) to $f:  \hhh(\kkk) \to
\hhh(\ppp)$ by linearity and introduce the border operator on
$\hhh(\ppp)$:

\be{54a}
\dd \ket{p} := \sum_{f(s)=p} \ket{f(\dd s)}
\ee

\noindent and verify the conditions (\ref{e49}). First calculate
its square:

\[
\dd^2 \ket{p} =
\sum_{f(s)=p} \dd \ket{f(\dd s)} =
\sum_{f(s)=p} \dd \sum\epsilon_{t_s}\ket{f(t_s)} =
\sum_s f(\ket{\dd \dd s}) = 0
\]

\noindent while the second condition (\ref{e49}) follows from
(\ref{54}) since for any component of the right hand side of the
sum

\[
\ketbra{\dd p}{q} =
\sum_{s:\, f(s)=p}\ketbra{f(\dd s)}{q}
\]

\noindent is in a unique block in accordance with (\ref{51}).

\medskip

So, the set $P$ of all proper elements of an arbitrary
atomic Greechie logic becomes discrete differential manifold.

\paragraph{Acknowledgement} The authors express their
gratitude to the participants of the Friedmann seminar
(St.Petersburg, Russia) for their patience and help in
eliminating numerous mistakes. We aknowledge the attention to
our work offered by prof. Richard Greechie.

The work was supported by the RFFI research grant and the 
research grant 97--14.3--62 "Universities of Russia".  
One of us (R.R.Z.) acknowledges the
financial support from the Soros foundation (grant a98-42).

\bigskip
\noindent{\bf\Large\bf References}
\medskip

Baehr, H.C., A.Dimakis and F.M\"uller-Hoissen, (1995),
Differential calculi on commutative algebras,
{\it Journal of Physics A: Mathematical and General},
{\bf 28},
3197
\smallskip

Birkhoff, G., (1967)
{\it Lattice Theory},
Providence, Rhode Island
\smallskip

Dimakis, A., and F.M\"uller-Hoissen, (1998),
Discrete Riemannian geometry,
eprint gr-qc/9808023
\smallskip

Djemai, A.E.F., (1995),
Introduction to Dubois-Violette's
Noncommutative Differential Geometry
{\it International Journal of Theoretical Physics},
{\bf 34},
801
\smallskip

Koszul, J.L., (1960)
{\it Fibre bundles and differential geometry},
Tata Institute of Fundamental Research, Bombay
\smallskip

Rota, G.-C., (1968),
On The Foundation Of Combinatorial Theory, I. The Theory Of
M\"obius Functions,{\it Zetschrift f\"ur
Wahrscheinlichkeitstheorie,\/}
{\bf 2},
340
\smallskip

Zapatrin, R.R., (1997),
Polyhedral representation of discrete differential manifolds,
{\it Journal of Mathematical Physics},
{\bf 38},
2741
(eprint dg-ga/9602010)
\smallskip

R.R.Zapatrin, (1998),
Finitary Algebraic Superspace,
{\it International Journal of Theoretical Physics},
{\bf 37}, 2, 799
eprint gr-qc/9704062

\end{document}